\documentstyle[11pt,aaspp4]{article}
\begin{document}
\title{SNI: a new mechanism for gamma ray burst. I.Weak magnetic field}
\author{V.S. Berezinsky}
\affil{INFN, Laboratori Nazionali del Gran Sasso\\
Statale 17bis 67010-Assergi (L'Aquila), Italy}
\author{P. Blasi}
\affil{INFN, Laboratori Nazionali del Gran Sasso\\ Universit\'a degli Studi
di L'Aquila\\
Via Vetoio, 67100-Coppito (L'Aquila), Italy}
\and
\author{B.I. Hnatyk}
\affil{Institute for Applied problems in Mechanics and Mathematics\\ NASU,
Naukova str. 3b, Lviv-53, 290053, Ukraine}
\begin{abstract}
We propose a new mechanism for high energy gamma ray burst in Supernova type I
(SNI) explosion.
Presupernova is assumed to be a binary system
comprised of a red giant and a white dwarf
with a wind accretion. The accretion flow terminates by the accretion shock
in the vicinity of the white dwarf at the distance of the order of the
accretion
radius. The gas inside the accretion radius constitutes the main body of a
target for gamma ray production.\par
The supernova explosion and the shock propagation in the white dwarf result in
the hydrodynamical acceleration of the outer layers of the star.
It proceeds in two stages. The first stage is caused by the shock propagating
in the outer layers of the star; the second stage is connected with the
adiabatic
expansion of the ejected shell into low-density medium around the white dwarf.
The spectrum of accelerated particles is steep and the maximum energy does not
exceed $1000$ $GeV$.\par
The gamma ray burst is produced due to interaction of the accelerated particles
with the gas in the binary system. Most of the photons have energies about
$100$ $MeV$, the total number of the emitted photons is between $10^{46}$ and
$10^{47}$. The typical duration of the burst is $\sim$ $1-3$ $s$ for
$\sim 100$ $MeV$ photons and $10^{-3}$ $s$ for $\sim 1$ $GeV$ photons.\par
The gamma ray burst might have one or two precursors. The first
one is produced during the shock breakout when the shock approaches the star
surface and crosses it.
This burst is produced by the heated gas behind the shock; the
radiation is blue-shifted due to the relativistic motion of the shell. The
second burst might be produced under the appropriate choice of the parameters
at the stage of the adiabatic expansion of the shell of the accelerated matter,
when the shell becomes transparent for radiation.\par
   The calculations are valid in the case of a weak magnetic field. The case of
strong magnetic field will be considered in part II (in preparation).
\end{abstract}


\section{Introduction}
It could be that there are a few different types of sources of observed gamma
ray bursts (GRB). If GRB have cosmological origin,the repeaters are sources
of a different type. GRB with smooth and complex time profiles can be
produced by two different types of sources. Probably burst radiation is a
typical phenomenon for many types of sources.\par
Our aim is not to explain the observed GRB but to search for reliable
production mechanisms, which can be discovered in future.
In this sense our paper is similar to the one by Berezinsky and Prilutsky
(1985)
\markcite{bere}
where the mechanism of GRB due to $\nu {\bar \nu}$ annihilation at the collapse
of a compact object was suggested. At that time this mechanism could not
explain the observed GRB, but it was used later for explanation of GRB in the
case of their cosmological origin.\par
We propose a mechanism of GRB in SNI explosion. More specifically, we
study
SNIa explosions, which as widely accepted occur in a binary system comprised
of a carbon-oxigen white dwarf with nearly Chandrasekhar mass and a late-type
secondary companion star (Wheeler and Harkness 1990, Khokhlov {\it et al.}
1993)
\markcite{Wheeler90}\markcite{Khokhlov93}.\par
The explosion is caused by the thermonuclear reactions due to disc accretion (
Nomoto 1982)\markcite{No92} or wind accretion (Munari and Renzini 1992, Kenyon
{\it et al.} 1993)\markcite{Mr92} \markcite{Klmt93} of the gas onto the white
dwarf. The shock wave propagation through the outermost layer of the white
dwarf accelerates a small part of it to relativistic and ultrarelativistic
velocities as it was first suggested by
Colgate and Johnson 1960 \markcite{colgjohn60}.\par
This acceleration is connected with the increase of the
shock velocity, when the shock propagates in a medium with
decreasing density.\par
For a long time the Colgate mechanism (see Colgate (1984)\markcite{Colg84}
 for a review) was one of the challengers for acceleration of the observed
cosmic rays. Now there are convincing arguments based on adiabatic energy
losses and on the estimates of rate of accelerated
particles, that this is not the case
(e.g. Woosley, Taam and Weaver 1986)\markcite{Wtw86}.\par \noindent
It is most probably true that the Colgate mechanism cannot provide all
observed pool of cosmic rays, but it is also true that this mechanism does work
under appropriate physical conditions. In particular it works in the
case of shock
propagation through the surface of a white dwarf.
The interaction of accelerated
particles with ambient gas results in the production of gamma radiation through
$p-p$ collisions, with the consequent $\pi^0$ production and decays.\par
In this paper we investigate a gamma ray burst produced due to the interaction
of accelerated particles with the accreting gas in the regime of wind accretion
. In Section 2 the space distribution of the gas in accretion flow is
considered. The propagation of the shock in the envelope of the white dwarf is
studied in Section 3. In Section 4 we
consider the additional acceleration of particles due to
expansion of the shell in the low-density medium and calculate the resulting
energy spectrum. The main gamma ray burst, produced as a result of the
interaction of accelerated particles with the accretion flow, is described in
Section 5. Discussion and conclusions are presented in Section 6.\par
\section{Gas distribution in presupernova binary system}
We shall concentrate mostly on the Supernova type Ia for which explosion occurs
in the white dwarf.
As a model for a presupernova  we shall consider a binary system comprised of a
carbon-oxygen white dwarf and a red giant. The accretion to the white dwarf
results in an explosion which occurs when the accreting
gas is accumulated on the carbon-oxygen interior of the compact companion and
triggers the detonation with consequent mass ejection.\par
   We shall consider the wind regime of accretion when the white dwarf accretes
the gas from the companion's stellar wind (Kenyon {\it et al.} 1993, Munari
and Renzini 1992)  \markcite{Klmt93} \markcite{Mr92}. In this regime
 of accretion an explosion is possible
for a carbon-oxygen  white dwarf with initial mass $0.5-1.3$
${M_{\odot}}$ and for
accretion rate of helium gas ${\dot M_a} \leq 4\times 10^{-8}$ ${M_{\odot}}/
yr$ (Fujimoto \& Taam 1982, Woosley \& Weaver 1993).
\markcite{fujitaam82}\markcite{Wtw86} \markcite{Ww93}.\par
   Observationally, such systems correspond to the symbiotic stars. The
observed orbital periods for these stars ($1-10$ $yr$) imply a separation of
the binary components $d\simeq (2-9)\times 10^{13}$ $cm$.
The secondary component generally does
not fill its Roche lobe and the mass loss is provided by the
stellar wind at the of
rate ${\dot M_W} \sim 10^{-7}$ ${M_{\odot}}/yr$ (Kenyon {\it et al.} 1993)
\markcite{Klmt93}.\par
As a first approximation, the accretion from the stellar wind can be described
by the
classical theory of Hoyle and Littleton (1939)\markcite{hoylit39}.
The parameters
of this theory are the mass of the accreting star $M_S$,
the density of the gas in the wind $\rho_W$ and its velocity in respect to the
star $u_W$. The theory operates with the accretion radius $R_{acc}$ and the
accretion rate ${\dot M_{HL}}$ given by:
\begin{equation}
R_{acc}={2 G M_S\over u_W^2 (d)}
\label{eq:racc}
\end{equation}
\noindent
and:
\begin{equation}
{\dot M_{HL}} = \pi R_{acc}^2 \rho_W (d) u_W (d)
\label{eq:mhl}
\end{equation}
\noindent
where $G$ is the gravitational constant and $d$ is the separation of the stars.
We shall assume the supersonic flow of the gas at the large distance from the
star $r \sim d$.\par
   The recent numerical calculations for two-dimensional axisymmetric wind
accretion were performed by
Matsuda, Inoue and Sawada (1987),\markcite{Mis87}
Fryxell and Taam (1988)\markcite{Fryxtaam88},
Taam and Fryxel (1988,1989)\markcite{Tf88} \markcite{Tf89},
Taam, Fu and Fryxell (1991)\markcite{Tff91}.
The hydrodynamical picture of the gas flow  includes
the shock, the region of stagnation $u_{acc}\simeq 0$, a considerable density
increase in the wake and some other features. The flow critically depends on
the accretion rate ${\dot M_a}$. For small accretion rates ${\dot M_a}<<
{\dot M_{Edd}}$ the flow is steady. For the large ${\dot M_a}$ it probably
exibits a time variation.
The details, essential for our calculations, are as follows. The shock is
produced at distance $R_{sh} = \xi R_{acc}$ with $\xi<1$ and typically $\xi=0.1
-0.2$. The accretion rate is ${\dot M_a}= \eta {\dot M_{HL}}$ where $\eta$ can
reach the large fraction $\eta\sim 0.6$. \par
   Inspired by these results we adopt the following simplified
picture for the binary system
 and for the gas flow in it (Fig. 1).\par
\placefigure{fig_1}
The compact star (white dwarf) and the secondary component (red giant) are
separated by distance $d$. The red giant does not fill its Roche lobe and its
mass loss is due to stellar wind with supersonic velocity $u_W$.
For our calculations it is enough to
assume the spherically-symmetrical shock with radius:
\begin{equation}
R_{sh} = \xi R_{acc}.
\label{eq:rsh}
\end{equation}
\noindent
Outside the shock the density distribution is:
\begin{equation}
\rho_W (r_{RG}) = {{\dot M_W}\over 4 \pi u_W r_{RG}^2}
\label{eq:redwind}
\end{equation}
\noindent
where $r_{RG}$ is the distance from the red giant. Inside the shock we assume:
\begin{equation}
\rho_a (r) = \rho_a (R_{sh}) (r/R_{sh})^{-3/2}
\label{eq:accre_den}
\end{equation}
\noindent
according to the calculations of Taam, Fu and Fryxel (1991)
\markcite{Tff91}.
Eq. (\ref{eq:accre_den}) corresponds to the quasi-spherical accretion inside
the shock with infall velocity proportional to the free-fall velocity:
\begin{equation}
u_a (r) = \zeta u_{ff} (r),
\label{eq:velocity}
\end{equation}
\noindent
where $u_{ff}=(2 G M_{S}/r)^{1/2}$ and $\zeta=\eta/(16\xi^{3/2})$.
In our calculations we shall use
$\eta=0.5$, $\xi=0.2$, and thus $\zeta=0.35$.\par
We assume the density jump on the surface of the shock as in the strong shock:
\begin{equation}
\rho_a (R_{sh}) = 4 \rho_W (d)
\label{eq:rhoa}
\end{equation}
\noindent
The numerical values which we use in the calculations are presented in Table
1. We shall summarize the notation used in Table 1: $R_S$ and $M_S$ are the
radius and the mass of the white dwarf, respectively, $\rho_0$ is given by
Eq. (\ref{eq:density}), $d$ is the separation between the stars in the binary,
$R_{sh}$ is the radius of the accretion shock, ${\dot M_a}$ is the rate of
accretion onto the white dwarf, ${\dot M_W}$ is the rate of mass-loss by the
secondary component of the binary, $u_W$ is the wind velocity and
$x_{sh}=\int_{R_S}^{R_{sh}} \rho(r) dr$ is the column density of
accreting gas inside the shock.\par
\placetable{table_1}

\section{The relativistic shock in SNI and the precursor gamma burst}
The explosive nuclear burning in the white dwarf produces an
outward-moving shock.
 In this section we shall study the propagation of the shock in the outer
layers of the white dwarf and beyond.\par
The process of thermonuclear burning propagates in the compact star either
in the form of a supersonic detonation wave with
velocity $(1.1-1.4)\times 10^9$
$cm/s$, or in the regime of a subsonic deflagration wave (Khokhlov,
M\"{u}ller \& H\"{o}flich 1993) \markcite{Kmh93}.The latter can be
converted into a detonation wave in the outer layers of the star where the
pressure gradient is high. We shall limit our consideration
in the propagation of the shock in the periphery of the star, where the
typical shock velocity is $u_{sh}\geq 10^9$ $cm/s$. The density of the gas in
the considered outer layer is characterized by a large gradient. In our
estimates we shall use the polytropic model with $n=3$ for which the density is
\begin{equation}
\rho(r)=\rho_0 (1-{r\over R_S})^3
\label{eq:density}
\end{equation}
\noindent
where $\rho_0$ is a constant and $R_S$ is the radius of the star.
The propagation of the shock in the outermost layers of the
white dwarf with large
density gradient results in the growth of the shock velocity up to the
relativistic regime (Colgate and Johnson 1960) \markcite{colgjohn60}.
The velocity of the strong adiabatic shock wave propagation in the medium
with large density gradient $-d ln(\rho)/d ln(r) >> 1$ in both the
nonrelativistic and relativistic regimes may be approximated by
the expression given by Gnatyk (1985) \markcite{Gna85}:
\begin{equation}
\Gamma_{sh}\beta_{sh} = const\times \rho^{-a},
\label{eq:gshbsh}
\end{equation}
\noindent
where $\beta_{sh}=u_{sh}/c$ is the dimensionless velocity of the shock, $c$
is the velocity of light and $\Gamma_{sh}=(1-\beta^2_{sh})^{-1/2}$ is the
Lorentz factor of the shock. The parameter $a$ in this
approximation is $a \approx 0.2$
for both nonrelativistic and relativistic shock waves. Analytical
solutions for the nonrelativistic shock wave  propagation in exponential and
polytropic density distribution give the values $0.17 \leq a \leq 0.23$
for abiabatic index $4/3 \leq \gamma_{ad}
\leq 5/3$ and polytropic index $1.5 \leq n \leq 3.25$. For the
ultrarelativistic shock Johnson and McKee (1971) \markcite{Johnmck71}
obtained  analytically
$a=0.232$. In our calculation we shall consider two extreme possibilities:
$a=0.2$ and $a=0.232$. \par
   Let us calculate now the maximum value of $\Gamma_{sh} \beta_{sh}$
for the shock in the star.
Following Colgate (1984) we shall use instead of $r$ another variable $F$,
which is the fraction of the star mass beyond radius $r$.
  Using eq.(\ref{eq:density}) one obtains:
\begin{equation}
F=\rho^{4/3} {\pi R_S^3 \over M_S \rho_0^{1/3}}.
\label{eq:fraction}
\end{equation}
\noindent
Then introducing the initial condition $\Gamma_{sh} \beta_{sh}=
(\Gamma_{sh} \beta_{sh})_i$ at $F =F_i$ we obtain for the shock velocity
in the shell:
\begin{equation}
\Gamma_{sh} \beta_{sh} = (\Gamma_{sh} \beta_{sh})_i ({F \over F_i})^
{-{3\over 4}a}.
\label{eq:gammabeta}
\end{equation}
   The numerical calculations (Khokhlov, M\"{u}ller and H\"{o}flich 1993 and
reference therein) \markcite{Kmh93} show that the shock starts to be
accelerated effectively
at the depth $F_i\sim 0.05$. This is caused by the large density gradient in
the outer layers of the white dwarf.
Hereafter we shall use $F_i= 0.05$ as the initial value.
According to the above mentioned numerical calculations the velocity of the
shock at $F=F_i$ is
\begin{equation}
(\beta_{sh})_i = 5\times 10^{-2} k_s,
\label{eq:betai}
\end{equation}
\noindent
with $k_s$ limited between $0.7$ and $1.0$. We shall use for our calculations
$k_s=1.0$.\par
The maximum value of $\Gamma_{sh}\beta_{sh}$ corresponds to the smallest
value of $F$ at which the shock still exists in the outer highly rarefied
layer. The shock breaks out at the radius $r_{max}$ where the dissipation of
energy due to the escape of photons or other particles
becomes essential. In the
case of radiation-dominated nonrelativistic SNII shock, this condition can be
expressed as an equality between the optical thickness of photons and the front
thickness (Imshennik and Nadyozin 1989, Ensman and Burrows 1992)
\markcite{Imshnad89}\markcite{Ensbur92}. In the case of relativistic SN shock,
the condition for the shock break out  can be taken as the equality of
the pathlenght of the nucleus in the shock
transition zone and the residual thickness of this layer
(Colgate 1974, 1984) \markcite{Colg74}\markcite{Colg84}
Using (\ref{eq:density}) and (\ref{eq:fraction}) one obtains:
\begin{equation}
F_{min} = {4\pi R_S^2 x_{int}\over M_S}
\label{eq:fmin}
\end{equation}
\noindent
where $x_{int}$ is the interaction pathlength of nuclei in $g/cm^2$.
Substituting this value of $F_{min}$ into Eq.(\ref {eq:gammabeta}) one obtains:
\begin{equation}
(\Gamma_{sh}\beta_{sh})_{max}=(\Gamma_{sh}\beta_{sh})_{i}
({4\pi R_S^2 x_{int}\over M_S F_i})^{-{3\over 4}a}.
\label{eq:gbmax}
\end{equation}

The value of $x_{int}$ depends on the physical conditions in the shock
transition zone. For the relativistic shock waves considered here, the main
source of shock dissipation is ion-lepton dynamical friction (Colgate 1974,
Weaver 1976)\markcite{Colg74}\markcite{Weav76}.
In the compact white dwarf case, the
temperature at the shock front nearby the surface is of the
order of $(3-4)\times 10^8 K$
(see below) and  the number density of electron-positron pairs reaches
$10^{23}-10^{24}$ $cm^{-3}$. The main energy losses of ions in this case
are due to elastic nucleus-electron scattering and bremstrahlung radiation.
The pathlength of ions is $1-3$ $g/cm^2$. In the extended white dwarf case the
temperature at the surface is lower and the energy losses are dominated by pion
production. However, deeper in the star at $x\sim 10$ $g/cm^2$, the
temperature is again high enough for $e^+ e^-$ production and the effective
pathlength can be taken as $x_{int}\sim 10$ $g/cm^2$.
Fortunatelly
according to eq. (\ref{eq:gbmax}) $(\Gamma_{sh}\beta_{sh})_{max}$ weakly
depends on $x_{int}$, typically as $x_{int}^{-0.15}$.
In the calculations below we use $x_{int}=3$ $g/cm^2$.
Eqs.(\ref {eq:gammabeta}) and (\ref {eq:gbmax}) describe the propagation of
the shock through the outer layers of the star. Namely, eq.(\ref
{eq:gammabeta})
gives the value of $\Gamma \beta$ for the shock at the various positions $F$
and eq.(\ref {eq:gbmax}) gives the maximum value of $\Gamma \beta$ , at the
moment when the shock reaches the surface of the star.\par
Now we shall proceed
with the calculations of the velocity of the gas and its thermodynamical
characteristics behind the shock.
   After the shock crosses a layer of the gas, this layer acquires a radial
velocity which is somewhat smaller than the shock velocity.
The gas behind the shock is also heated to a high temperature. This is the
first
stage of hydrodynamical acceleration.\par
   Let us give the typical values of $(\Gamma_{sh}\beta_{sh})_{max}$ in our
models. For the  {\it compact white dwarf} model with $M_S=1.4M_{\odot}$,
$R_S=1\times 10^8 cm$ and $\rho_o = 1\times 10^{12} g/cm^3$ (Table 2),
$F_{min}$ from eq.(\ref {eq:fmin}) is $4.5\times 10^{-17}$ and hence
$(\Gamma_{sh}\beta_{sh})_{max} = 9.0$ and $20$ for $a=0.2$ and $a=0.232$,
respectively. For the {\it extended white dwarf} model with $M_S=
1.0M_{\odot}$, $R_S=1\times 10^9 cm$ and $\rho_o = 4\times 10^9 g/cm^3$
(Table 1) we obtain $F_{min}=6.3\times 10^{-15}$ and
$(\Gamma_{sh}\beta_{sh})_{max} = 4.3$ and $8.8$ for $a=0.2$ and $a=0.232$,
respectively. Note that these values are considerably smaller than the ones
in the calculations of  Colgate (1984) \markcite{Colg84}.\par
\placetable{table_2}
   Heating and acceleration of the gas behind the strong adiabatic
radiation-dominated ($\gamma_{ad}=4/3$) shock wave propagating in
nonrelativistic cold gas can be described in terms of the conditions on the
shock front as given by Blandford and McKee (1976) \markcite{Blmk76}:
\begin{equation}
\epsilon_2 = {\Gamma_2} (4 {\Gamma_2} +3) \rho_1 c^2
\label{eq:epsilon2}
\end{equation}
\begin{equation}
\rho_2 = (4 {\Gamma_2} +3) \rho_1
\label{eq:rho2}
\end{equation}
\begin{equation}
\Gamma^2_{sh} = (4 {\Gamma_2} - 1)^2 ({\Gamma_2} + 1)/(8{\Gamma_2} + 10)
\label {eq:gsggas}
\end{equation}
\noindent
where indicies $1$ and $2$ refer to the parameters in front and behind the
shock, respectively,
${\Gamma_2} = (1-\beta_2^2)^{-{1/2}}$ is the Lorentz factor of the
fluid behind the shock, $\rho$ is the density of the gas and  $\epsilon$ is
the energy density.\par
The equation of state of the gas behind the shock is given by:
\begin{equation}
p_2 = {1\over 3} (\epsilon_2 - \rho_2 c^2)
\label{eq:pressure}
\end{equation}
\noindent
{}From eq. (\ref{eq:pressure}) it is easy to obtain the following formula for
the
pressure:
\begin{equation}
p_2 = {1\over 3} (4 {\Gamma_2} + 3) ({\Gamma_2} - 1) \rho_1 c^2.
\label{eq:pressure_1}
\end{equation}
\noindent
   In our case the main contribution to the pressure is provided by the
radiation
\begin{equation}
p_2 = {1\over 3} a_K T^4_2,
\end{equation}
\noindent
for $kT_2 << m_ec^2$, and by the radiation and the electron-positron pairs
\begin{equation}
p_2 = {1\over 3} a_K T^4_2 (1+ {7\over 4})
\label{eq:pressure_2}
\end{equation}
\noindent
for $kT_2>>m_e c^2$, where $a_K$ is a constant of the energy density of
radiation.\par
Hence one obtains for $kT_2>>m_e c^2$ and $kT_2<<m_e c^2$ respectively:
\begin{equation}
T_2 = ({4\over 11 a_K}(4 \Gamma_2 + 3)(\Gamma_2 - 1)\rho_1 c^2)^{1/4}
\label{eq:t2_1}
\end{equation}
\noindent
and
\begin{equation}
T_2 = ((4 \Gamma_2 + 3)(\Gamma_2 - 1) {\rho_1 c^2\over a_K})^{1/4}
\label{eq:t2_2}
\end{equation}
  A comment to eq. (\ref {eq:gsggas}) is in order. This equation gives the
connection between the Lorentz factor of the shock $\Gamma_{sh}$ and the
Lorentz
factor $\Gamma_2$ of the fluid behind the shock. Therefore, from the known
shock velocity at each $F$, given by Eq. (\ref {eq:gammabeta}), one can
reconstruct using Eq.(\ref {eq:gsggas}) the distribution of Lorentz factor
of the gas over the outer layers at the moment when the shock reached the
surface of the star. The maximum value of $\Gamma_2$ corresponds to maximum
value of $\Gamma_{sh}$ given by eq. (\ref {eq:gbmax}). Numerically for the
{\it  compact white dwarf} model we obtain $(\Gamma_2)_{max} = 6.7$ and
$15$ for $a=0.2$ and $a=0.232$, respectively. For the {\it extended white
dwarf} model $(\Gamma_2)_{max} = 3.4$ and $6.6$ for $a=0.2$ and $a=0.232$,
respectively. According to eq. (\ref{eq:t2_1}), the large value
of $(\Gamma_2)_{max}$ results in the high temperature $T_2^{sur}$
at the moment when the shock reaches the surface.\par

   Let us discuss now the temperature of the gas behind the shock front and
the precursor gamma burst connected with it. When the shock reaches the
surface of the star, the temperature $T_2^{sur}$ behind it is given by
Eq. (\ref{eq:t2_1}) with
$\Gamma_2=(\Gamma_2)_{max}$. The corresponding values for the {\it
compact} model are $4.9\times 10^8 (7.4\times 10^8)K$ for $a=0.2(0.232)$ and
for the {\it extended} model $1.6\times 10^8(2.2\times 10^8)K$ for
$a=0.2(0.232)$.\par
The burst of radiation emerges from the star during the shock breakout,
when radiation energy trapped in the shock escapes through the trunslucent
external layer of the star. The shock breakout phenomenon was studied in
detail for the case of SNII by Grassberg, Imshennik and Nadyozhin (1971),
Imshennik and Nadyozhin (1988, 1989), Blinnikov and Nadyozhin (1991), Ensman
and Burrows (1992)
\markcite{Gin71}\markcite{Imshnad88}\markcite{Imshnad89}\markcite{Bn91}
\markcite{Ensbur92}.
The picture can be described as follows.
At the time the shock approaches the surface
it becomes thick. When the precursor of the shock reaches the surface, the
front is spread over the distance with the optical thickness
$\tau \sim 100$, according to the calculations of Ensman and Burrows (1992)
\markcite{Ensbur92}.
This moment can be taken as one  when the shock breakout starts and
the radiation emerges from the star. Following qualitatively this picture
it is easy to evaluate the total energy of the burst in the relativistic
case that we are examining.
In the rest system of the shock, this is given by the product of
$4\pi R_s^2 (T_2^{sur})^4$ and the duration of the burst. The latter can
be estimated as the time, $\delta t_{rest} \sim l_{\gamma}/c$, of the photon
escape from the thickness, $l_{\gamma}$, of the shock front. For the
{\em compact dwarf}, scattering of the photons is caused mainly by
$e^+e^-$-pairs (for the {\em extended dwarf} their contribution is rather
small). Then $l_{\gamma}$ for optical thickness $\tau \sim 100$ is given by
$l_{\gamma} \sim \tau (\sigma_T n_{e^+e^-})^{-1}\sim 2.5\cdot 10^2 (2.5)
{}~cm$ for $a=0.2 (0.232)$. Taking into account the blue shift of the emitted
photons, the fluency $F_{\gamma}$ can be found as
\[
F_{\gamma}=(\Gamma_2)_{max}
4 \pi R_s^2\sigma (T_2^{sur})^4l_{\gamma}/(4\pi r^2 c)
\]
where $r$ is the distance to the source.  Numerically this fluency is
$F_{\gamma} \sim 2 \times 10^{-6}~erg/cm^2$ for $r= 10~kpc$ and the compact
dwarf model with $a=0.2$. For the {\em extended white dwarf model} this
flux is $\sim 10^2$ times larger.
The typical energy of the photons in the burst is
$E_{\gamma}^{obs} \sim (\Gamma_2)_{max} k T_2^{sur} \sim 300 (1000)~keV$
for the case of {\em compact white dwarf} with $a= 0.2 (0.232)$. The values
of the temperature $T_2^{sur}$ and the mean energy of the photons from the
precursor burst are tabulated in Table 2.
The duration of the burst, $\delta t_{obs}$, as observed by a distant
observer is very short. It is determined by the time delay of the photons
arriving from the different points of the star surface. The beaming effect
makes the duration shorter in the relativistic case (see e.g. Rees 1968).
Since the emitting angle is $\theta_e \sim 1/\Gamma$ in the laboratory frame,
the time delay of the photons from the $\theta=0$ and
$\theta = \theta_e$ directions is
\begin{equation}
(\delta t)_{obs} \sim \frac{R_s}{c}\frac{1}{2\Gamma_2^2}
\label{eq:delta_t}
\end{equation}
\noindent
which numerically is $4\times 10^{-5} (7\times 10^{-6})~s$ for
$a=0.2 (0.232)$.

The accurate calculations are needed for the detailed description. However,
the gamma ray precursor due to the shock breakout seems to be detectable by
such instruments as BATSE.

\section{Hydrodynamical acceleration}

Hydrodynamical acceleration proceeds in two stages. During the first one,
described in section $3$, the shock propagating through the gas with the
falling density, increases its velocity according to eq.(\ref{eq:gammabeta})
until the fraction
of the shell ahead of it reaches $F_{min}$ given by eq. (\ref{eq:fmin}).
Starting from this
moment the shock is decaying. It leaves behind the heated shell moving
outwards with velocity $\beta_2$ and Lorentz factor
$\Gamma_2$ given by eq. (\ref{eq:gsggas}). The shell now expands
in the low
density gas. The thermal energy of the shell as well as the work of the
pressure of
the inner layers, are converted into kinetic energy of the shell. This is the
second, adiabatic, stage of acceleration.\par
   When the shock reaches the surface of the star, the shell has a wide
velocity distribution from nonrelativistic values in the inner
parts of the shell to ultrarelativistic ones in the narrow outer part
of the shell. During the adiabatic stage, the nonrelativistic
velocities  of gas elements increases as $u_f \simeq 2 k_N u_2$, where
$u_f$ and $u_2$ are the final and initial velocities, respectively, and
we take $k_N \sim 1$ ( $k_N=0.91$ for $n=3$ polytrope and $\gamma_{ad}=4/3$
(Kazhdan and Murzina 1992)). The initial velocity $u_2$ is given by Eq.
(\ref{eq:betai}).
In the ultrarelativistic case

\begin{equation}
\Gamma_f = \Gamma_2^b.
\end{equation}

The power $b$ as obtained in the analytical
solutions  is $b=1+{\sqrt 3}=2.73$ for the
plane shock wave (Johnson \& McKee 1971) \markcite{Johnmck71}
and $b=2.0$ for the spherical one
(Eltgroth 1972) \markcite{elt72}.\par
In our calculations we shall adopt $2.0 \leq b \leq 2.73$.\par
One can approximate both relativistic and nonrelativistic cases by the
relation:

\begin{equation}
\Gamma_f = \Gamma_2^b + (4 k_N^2 -b) {\Gamma_2 - 1 \over \Gamma_2^2}
\label{eq:gammaeffe}
\end{equation}
\noindent

Therefore,
the gas element located in the pre-shock shell at a distance $r$, which we
describe by the fraction of external
mass $F$ given by eq. (\ref{eq:fraction}), is accelerated
first to the Lorentz factor $\Gamma_2 (F)$
, given by eqs. (\ref{eq:gammabeta}),(\ref{eq:gsggas})
and then to the Lorentz factor $\Gamma_f$ given by eq.
(\ref{eq:gammaeffe}).\par
The number of nucleons accelerated to a kinetic energy per nucleon higher
than $E_K=(\Gamma_f - 1)m_H c^2$ is given by

\begin{equation}
Q_N(>E_K)={(F(E_K) - F_{min}) M_s \over m_H}.
\end{equation}

   Therefore, $Q_N(>E_K)$ gives us the integral spectrum of hydrodynamically
accelerated nuclei. \par
For the nonrelativistic case $\Gamma_f -1 << 1$ we have:
\begin{equation}
F(E_K)=F_i (E_K/E_i)^{-\gamma_{nr}^{int}}
\label{eq:fek}
\end{equation}
\noindent
where $\gamma_{nr}^{int}=2/(3a)$ and numerically $\gamma_{nr}^{int}=3.3 (2.9)$
 for $a=0.2 (0.232)$. The energy parameter $E_i$ is given by:
\begin{equation}
E_i = {72\over 49} (\Gamma\beta)_i^2 m_H c^2 = 3.4 MeV/nucleon.
\end{equation}
For ultrarelativistic case $\Gamma_f >> 1$ one obtains
\begin{equation}
F(E)=
K({E\over m_H c^2})^{-\gamma_{rl}^{int}}
\label{eq:fe}
\end{equation}
\noindent
where $K=1.1\times 10^{-11}$ and $K=2.3\times 10^{-10}$ for $a=0.2$ and
$a=0.232$, respectively. The exponent of the spectrum in ultrarelativistic case
is $\gamma_{rl}^{int}=4/(3ab)$ and numerically it is equal to
$\gamma_{rl}^{int}=3.3$ for
$a=0.2$ and $b=2.0$; and $\gamma_{rl}^{int}=2.1$ for $a=0.232$ and $b=2.73$
\par\noindent
Note that in all cases the values of $\gamma$ are given for integral spectra.
\par
The parameters of the spectra of hydrodynamically accelerated particles
are given in Table $3$.
\placetable{table_3}
\par
Table $3$ shows that in the {\it compact white dwarf} case the Lorentz factor
of hydrodynamically accelerated nuclei
reaches the values of $\Gamma_f \sim 100-1000$, while for the
{\it extended white dwarf} the
maximum values of Lorentz factors are less, $\Gamma_f \sim (10-100)$. The
fraction $10^{-9}-10^{-10}$ of the
total mass of the star is accelerated to the Lorentz factors
$\Gamma_f \geq 2$. The net energy of these particles reaches
$10^{44}-10^{45}$ $ergs$. The net energy of the subrelativistic particles
with $E_K>500$ $MeV/nucleon$
is $10^{45}-10^{46}$ $ergs$. The total number of nucleons $Q_N$ accelerated to
energies higher than $E_K$ is given in Table 3.
 The particles with these energies can generate
gamma radiation through production of neutral pions. $E_{\gamma}^m$ in
Table 3 refers to the end of the power low spectrum.\par
During the expansion of the relativistic shell its optical depth $\tau$
decreases. When $\tau>>1$ the energy losses of the shell due to radiation are
small. The thermal energy of the shell and the work of the pressure of the
internal layers are converted into kinetic energy of the shell and these
processes are taken into account in the above mentioned calculations. Due to
the work
of the inner layers, the total energy of the outer layer in
the end of the adiabatic expansion is larger than its total energy at the
moment when the shock crossed it.\par
   When the optical depth of the shell decreases to $\tau\leq 1$, the remaining
thermal energy will be mostly radiated away and at some conditions a second
precursor gamma ray burst might be produced. The physics of
this second burst is similar to that of the prompt precursor of gamma burst
in the fireball models (Goodman 1986; Paczynsky 1986; Shemi and Piran
1990; Meszaros and Rees 1993)\markcite{Goodman86}\markcite{Pa86}
\markcite{Shpi90}\markcite{Mr93}.

\section{Gamma ray burst}
   The expansion of the relativistic shell after the adiabatic phase looks like
the propagation of a powerful beam of relativistic particles. They have
energies up to $1000$ $GeV/nucleon$ and total energy $10^{45}-10^{46}$
$ergs$.
These particles interact with the accreting gas producing the pions and
gamma rays from neutral pion decays.
The distribution of the gas around the white dwarf is described by
eq. (\ref{eq:redwind}) for
$r\geq R_{sh}$ and eq. (\ref{eq:accre_den}) for $r\leq R_{sh}$.
The numerical parameters for this distribution are listed in Table $1$.\par
   Our calculations for production of photons (through $pp \to \pi^0 \to
\gamma\gamma$) are performed for two energy regions:
$E_K^{min}\leq E_{K}\leq E_c$ and $E_{K}> E_c$, where $E_K^{min}=0.4 GeV/
nucleon$ and $E_c=10 GeV/nucleon$.\par
For the low energy region $0.4-10$ $GeV/nucleon$ we calculate the total number
of photons $Q_{\gamma}$ of all energies produced in the burst at angle $\theta$
relative to the direction to the secondary companion:

\begin{equation}
{d N_{\gamma} (\theta)\over d \Omega}=2
\int_{R_S}^{\infty} dr n_N (r,\theta) \int_{E_{K}^{min}}^{E_c}
d E_{K} {Q_N (E_{K})\over 4\pi} <\xi\sigma(E_{K})>
\label{eq:flux}
\end{equation}
\noindent
where $n_N (r,\theta)$ is the number density for the nucleons in the gas,
$Q_N (E_{K})$ is the total number of accelerated nucleons with energy
$E_{K}$ per nucleon and $<\xi\sigma(E_{K})>$ is the cross section for
$\pi^0$ production in $NN-$ collisions. For this cross-section we use the
parametrization of experimental data for $p+p \to \pi^o + all$ given by Dermer
(1986) \markcite{dermer86}.\par
For energies $E>10$ $GeV/nucleon$  we shall use the scaling approximation
for the cross section:

\begin{equation}
{d\sigma(E_N,E_{\pi})\over dE_{\pi}} = {\sigma_0 \over E_{\pi}} f_{\pi}(x)
\label{eq:cross}
\end{equation}
\noindent

where $x=E_{\pi}/E_K$, $E_{\pi}$ is the pion energy, $\sigma_0 = 32 mb$ and:

\begin{equation}
f_{\pi}(x)=0.67(1-x)^{3.5}+0.5 e^{-18x}
\label{eq:fpion}
\end{equation}
\noindent

This approximation describes very well the experimental data. The differential
spectrum of produced gamma radiation is calculated as

\begin{equation}
{dN_{\gamma}(E_{\gamma}, \theta)\over d\Omega dE_{\gamma}} = \sigma_0
\int_{R_S}^{\infty} dr n_N (r,\theta) \int_{E_{\gamma}}^{E^{max}_K}
{d E_{K}\over E_{K}}
{Q_N (E_{K})\over 4\pi} \int_{E_{\gamma}/E_{K}}^1 {dx\over x^2}
f_{\pi}(x)
\label{eq:flux_1}
\end{equation}
\noindent

where $E^{max}_K$ is the maximum energy per nucleon in the spectrum of
accelerated protons.
The results of the calculations are displayed in
Figs. $2$-$4$.\par

In Fig. $2$ the column density of gas in the binary system is plotted
as a function of angle $\theta$ relative to the line between the centers of
the stars. One can see that for all angles the main contribution comes from
the region $r\leq R_{sh}$ (the horizontal lines).
Only at small angles the contribution of the
stellar wind is appreciable, however for $\theta > 5^o$ it does not exceed
$10\%$.\par
\placefigure{fig_2}

   The column density in both our models is considerably less than the
nuclear interaction length ($x_N\sim 40-50 g/cm^2$) and high-energy gamma
absorption length ($x_{abs}\sim 70 g/cm^2$). Therefore, the column density
dependence $x(\theta)$ shown in the Fig. $2$, coincides exactly with the
total photon flux dependence $ \dot N_{\gamma}(\theta)$. \par

   More explicitly the distribution of the gas in the binary system is
displayed in Fig. $3$, where the density of the gas, $\rho$, is plotted against
the distance from the white dwarf. The position of the shock with the
transition to the stellar wind
density is clearly seen there.\par

  In Fig. $3$ the density profile coincides with the time profile of the
gamma burst. The energy-
integrated number of produced photons $d \dot N_{\gamma}/d\Omega$ for given
$\theta$ (angle relative to the binary axis) is shown as a function of the
production time $t$ in the binary system. This time is equal to the
time-of-flight, $t=(r-R_S)/v_N$, for the nuclei with energy per nucleon $E_K$
at the maximum of the production rate $<\xi \sigma (E_K)>N(E_K)$.
\placefigure{fig_3}
\par
   The flux decreases with time approximatively as $(t+t_o)^{-3/2}$, where
$t_o=
R_S/v_N$.
Duration of the burst for a remote observer is different at different
energies of photons.
The peak in the spectrum, centered at $E_{\gamma} \approx 70~MeV$ is produced
by pions with small kinetic energies and therefore photons are emitted
isotropically.
The observed duration is determined by difference in time-of-flight
for photons arriving radially ($\theta=0$) and from the edge of the limb
($\theta \sim \bar{R}/r$), where $\bar{R}=\sqrt{R_sR_{sh}}$ is the mean
length of the $pp$-interaction  region and $r$ is the distance between the
observer and the source.
It gives the duration $(\Delta t)_{obs} \sim 1~s (3~s)$
 for the {\em compact (extended)} model.
For the high energy photons produced by the protons with
$\Gamma \gg 1$ the burst
is considerably shorter. The simple kinematic estimates show that the photons
are on average emitted at an angle $\theta \approx 0.3/\Gamma$ relative to the
direction of the proton propagation (i.e. the radial direction). Then the
duration, determined again as the maximum difference of time-of-flight, is

\[
(\Delta t)_{obs}=\frac{\bar{R}}{c}\frac{\theta^2}{2} \approx
0.05 \frac{\bar{R}}{c}\Gamma^{-2}.
\]

In particular for photons with $E_{\gamma} \sim 1~GeV$ the observed
duration of the burst is $(\Delta t)_{obs}= 5\cdot10^{-4} (1\cdot 10^{-3})~s$
for the {\em compact (extended)} model.

\placefigure{fig_4}

The spectrum of the high energy gamma radiation is very steep: it has the same
power-exponent $\gamma_{rl}^{int}+1$ as the differential spectrum
of accelerated particles. At $E_{\gamma}\simeq 70$ $MeV$ it has the well-known
peak from the decays of pions at rest. Therefore, the
gamma burst is detected most effectively at $E_{\gamma}\sim 100 MeV$. However,
at $E_{\gamma}\geq 1 GeV$ the signal is also detectable.
The total number of emitted photons varies from $N_{\gamma}\sim 4.5 \times
10^{47}$ (for the {\it compact white dwarf  model}) to
$N_{\gamma}\sim 3.5 \times 10^{45}$ (for the {\it extended white dwarf
model}) (Table 4).
The corresponding fluences are in the range from
$0.3$ to $40$ $photons/cm^2$ for a distance of $10$ $kpc$.

\placetable{table_4}

\section{Discussion and Conclusions}

We study SNIa explosion in a binary system comprised of a compact star (white
dwarf) and a late-type secondary companion (red giant).
The accretion onto a white dwarf
causes the ignition of nuclear reactions and the ejection of the shell. The
accretion is assumed to be in the wind regime with the Roche lobe of the
secondary companion not filled. The accretion flow terminates by the shock at
distance $r=R_{sh}$ from the white dwarf. The radius of the shock $R_{sh}$ is
of the order of the accretion radius $R_{acc}$, see eq. (\ref{eq:racc}).
At $r\leq R_{sh}$ we assume a radial accretion flow to the white dwarf.\par
   Gas within the shock radius comprises the main fraction of the target for
gamma ray production. Its density and column density are determined mostly by
the mass of the compact star $M_S$ and by the parameters of the
stellar wind, its
density $\rho_W$ and velocity $u_W$. We take the density distribution and
the radius of the shock $R_{sh}$ according to the numerical
calculations by Taam,
Fu and Fryxell (1991) \markcite{Tff91}.  The values of the column density
of the gas as a function of
$\theta$, the angle relative to the line between the centers
of the two stars, are
given in Fig. 2. The mean value of the column density is $x=3.4$ $g/cm^2$ for
the {\it compact white dwarf} model and $x=0.67$ $g/cm^2$ for the {\it extended
white dwarf} model (Table 1).\par
The particles in our model are accelerated hydrodynamically. We distinguish two
stages of acceleration. The first one is connected with the propagation of
the shock through the shell. The acceleration occurs if the density in the
shell decreases sharply outwards: e.g. $\rho (r)\sim r^{-m}$ with $m>3$. In
this case the shock propagates with increasing velocity leaving the hot gas
with large radial velocity behind the shock front. Following Colgate (1984)
\markcite{colg84},
we describe the position of the considered layer by the amount of the gas in
front of it. Namely we use the fraction $F=M(>r)/M_s$, where $M(>r)$ is the
mass of the shell at the distance greater than $r$ and $M_s$ is the mass of the
star. The Lorentz factor of the gas in the shell at given $F$ can be found from
eqs. (\ref{eq:gammabeta}), (\ref{eq:gsggas}).
The uncertainties for these calculations are mainly connected with the
parameter $a$, which is defined by eq. (\ref{eq:gshbsh}); we expect
that it varies from $0.2$ to $0.232$ (Gnatyk 1985)
\markcite{Gna85}.\par
The second stage of acceleration begins when the shock reaches the outer edge
of the star. The shell adiabatically expands into a low-density medium and the
pressure accelerates the shell. The thermal energy of the gas and the work of
the pressure are converted into kinetic energy of the
shell. This stage terminates when the shell becomes transparent and the thermal
energy is radiated away.
The Lorentz factor of the shell acquired at
this stage is $\Gamma_f=\Gamma_2^b$, where $\Gamma_2$ is the Lorentz factor at
the end of the first stage and $b$ is a parameter which varies from $b=2.0$ for
the spherical wave (Eltgroth 1972) \markcite{Elt72}
to  $b=2.73$ for the plane wave (Johnson \& McKee 1971)\markcite{Johnmck71}.
The value of $b$ gives the largest uncertainties for the
calculations of this stage of acceleration.\par
   The numerical predictions are given for two models: {\it compact white
dwarf}
($M_S=1 M_{\odot}$ and $R_S=1 \times 10^8 cm$) and {\it extended white dwarf}
($M_S=1 M_{\odot}$ and $R_S=1 \times 10^9 cm$)(Table 1). The former case is
more
favorable for particle acceleration. For given values of the
parameters $a$ and
$b$ the spectrum of the accelerated nuclei has a power-law form
with the integral
exponent $\gamma_{rl}^{int}=4/(3ab)$. In the {\it compact white dwarf} case
with $a=0.232$, $b=2.73$ we obtain $\gamma_{rl}^{int}=2.1$, the maximum energy
of particles $E^{max}\sim 1400 GeV/nucleon$, and the total energy of particles
with $E_K>500 MeV/nucleon$ $W\sim 3\times 10^{46} ergs$. \par
 For the other extreme case of {\it extended white dwarf} and the values
$a=0.2$, $b=2.0$, the acceleration has a minimal efficiency:
$\gamma_{rl}^{int}=
3.3$, $E^{max}\sim 11$ $GeV/nucleon$ and the
total energy is $W\sim 2\times 10^{45} ergs$.\par
Two precursor gamma ray bursts are expected. The first one is connected with
the shock breakout when the shock traverses
the outer thin layer transparent for radiation. The
thermal radiation accumulated in the shock front is radiated away when the
shock reaches the surface of the star.
The thermal photons are
blue-shifted due to the relativistic expansion of the shell and their typical
energies reach $E\sim \Gamma_2 k T_2\sim 300(1000)$ $keV$ for the
$a=0.2(0.232)$
{\it compact} model and $E\sim 47(120)$ $keV$ for the {\it extended} one.
Relativistic effects probably convert the Planck spectrum in the power-law
spectrum (Shemi 1994)\markcite{Shemi94}.
   The total energy of the burst in the reference frame of the observer is
$W_{burst}\sim 2\times 10^{38}(2\times 10^{37})$ $ergs$ for the $a=0.2(0.232)$
{\it compact} model and $W_{burst}\sim 6\times 10^{40}(3\times 10^{41})$
$ergs$ for the {\it extended} one.
   The duration of the burst for a distant observer is given by
 eq.(\ref{eq:delta_t}). This time is typically very
short $\Delta t \sim 10^{-5}$ $s$ for a {\it compact} white dwarf.\par
   The second precursor burst might appear at the last stage of the adiabatic
expansion of the shell of accelerated particles behind the shock,
when this shell becomes transparent and its thermal energy is radiated
away.\par
   The main gamma burst is produced due to the interaction of the
accelerated nuclei
with the accreting gas. The major part of the calculated flux
corresponds to the interaction with the gas within the shock radius $R_{sh}$.
Protons are produced due to the neutral pion decays. At high energies
the gamma ray spectrum is very steep: it is
characterized by the same spectrum exponent
of the accelerated particles (see eq. (\ref{eq:fek}) and (\ref{eq:fe})).
At low energy the spectrum has the usual peak at
$E_{\gamma}\simeq 70$ $MeV$.
Most of the photons have energies $E_{\gamma}\geq
70-100$ $MeV$. The maximum energies of photons are of order $E_{\gamma}\sim
4-140$ $GeV$ depending on the values of the parameters $a$ and $b$ for the
{\it compact white dwarf} model and $E_{\gamma}\sim 1-15$ $GeV$ for
the {\it extended white dwarf} one. The total number of photons in the
burst is $N_{\gamma}^{tot} \sim  2\times10^{46}-5\times10^{47}$ for
the {\it compact white dwarf} and $N_{\gamma}^{tot} \sim 4\times10^{45}-
1\times10^{47}$ for the {\it extended white dwarf}. \par
   We want to emphasize here that in our calculations we used very
conservative choise of the parameters deliberately trying not to
overestimate the predictions. The accelerated particles traverse only
about $3.5 g/cm^2$ and $0.67 g/cm^2$ of the matter in {\it compact} and
{\it extended} models respectively.
For acceleration we used very conservative assumptions too, which
resulted in a steep spectrum and low maximum energies of the
accelerated particles
in comparison with the Colgate (1984) \markcite{Colg84} calculations. \par
   The predicted burst radiation can be easily detected if
SNIa occurs in our Galaxy. For a distance $10$ $kpc$ from the Earth, the
typical flux of gamma radiation at $E_{\gamma} \geq 100$ $MeV$ is $1$
$phot/cm^2$, for
$E_{\gamma} \geq 10$ $GeV$ it is $10^{-4}-10^{-3}$ $phot/cm^2$. \par
   The duration of the burst is different in different energy intervals.
At $E_{\gamma} \geq 100$ $MeV$ it is of the order of $1$ $s$.
For the high energy photons at $E_{\gamma} \geq 10$ $GeV$ it
reduces to $10^{-5}$ $s$.
Thus we predict a novel type of gamma ray bursts with properties quite
different from those presently observed. Since these bursts are rare
phenomenon, their detection needs the patrol service.\par
The hydrodynamical mechanism for gamma ray bursts which we put forward here
for SNI can work for other objects too. The crucial condition for the
hydrodynamical mechanism is a large density gradient which provides
acceleration of the shock front. The accelerated particles produce  a
detectable gamma ray burst even in the case of relatively small column
density of the target.
As an example we can mention the accretion-induced collapse of a white dwarf
to a neutron star (Woosley and Baron 1992 and references therein)
\markcite{Wb92}.
According to these calculations $7\cdot10^{-6}$ part of the whole dwarf
mass acquires velocity higher than $u_f=38000~km/s$. Then maximum velocity
of the shock is given by $(\Gamma_{sh}\beta_{sh})_{max}=2.8 (5.2)$
for $a=0.2 (0.232)$ which is only a factor 1.5 smaller than in the case
of our {\em extended} model. Therefore, we expect here the gamma burst
too.
Another example can be given probably by the explosion of a neutron star below
the minimum mass (Blinnikov et al 1990, Colpi, Shapiro and Teulkovsky 1993)
\markcite{Betal90}\markcite{Cst93}.
These examples illustrate the attractiveness of the hydrodynamical mechanism
for astrophysical applications.

\acknowledgments
B. Hnatyk is grateful to Laboratori Nazionali del Gran Sasso and in particular
to the director Prof. P. Monacelli for hospitality.

\newpage
\begin{table}
\caption{The values of the parameters for the white dwarf and the
wind accretion flow. \label{table_1}}
\begin{center}
\begin{tabular}{l c c }
\hline \hline
parameters & {\em Compact white dwarf} & {\em Extended white dwarf} \\ \hline
$R_S,cm$ & $1.0\cdot 10^{8}$ & $1.0\cdot 10^{9}$ \\
$M_S/{M_{\odot}}$ & $1.4$ & $1.0$ \\
$\rho_0,g/cm^3$ & $1.0\cdot 10^{12}$ & $4.0\cdot 10^{9}$ \\
$d,cm$ & $3.0\cdot 10^{13}$ & $3.0\cdot 10^{13}$ \\
$R_{sh},cm$ & $8.3\cdot 10^{12}$ & $5.9\cdot 10^{12}$ \\
${\dot M_a},{M_{\odot}}/yr$ & $2.4\cdot 10^{-8}$ & $1.2\cdot 10^{-8}$ \\
${\dot M_W},{M_{\odot}}/yr$ & $1.0\cdot 10^{-7}$ & $1.0\cdot 10^{-7}$ \\
$u_W,cm/s$ & $3.0\cdot 10^{6}$ & $3.0\cdot 10^{6}$ \\
$x (r\leq R_{sh}), g/cm^2$ & $3.4$ & $0.67$ \\
\hline

\end{tabular}
\end{center}

\caption{The values of the shock and gas parameters at the moment of shock wave
break out.
\label{table_2}}
\begin{center}
\begin{tabular}{l c c  c c }
\hline \hline
 & \multicolumn{2}{c}{\em Compact white dwarf} &
\multicolumn{2}{c} {\em Extended white dwarf} \\

Parameters & {$a=0.2$} & {$a=0.232$} & {$a=0.2$} & {$a=0.232$} \\ \hline

$F_{min}$ & $4.5\cdot 10^{-17}$ & $4.5\cdot 10^{-17}$ & $6.3\cdot 10^{-15}$ &
$6.3\cdot 10^{-15}$ \\
$(\beta_{sh}\Gamma_{sh})_{max}$ & $9.0$ & $21$ & $4.3$ & $8.8$ \\
$(\Gamma_2)_{max}$ & $6.7$ & $16$ & $3.4$ & $6.6$ \\
$T_2^{sur}, K$ & $4.8\cdot 10^{8}$ & $7.4\cdot 10^{8}$ & $1.6\cdot
10^{8}$ & $2.2\cdot 10^{8}$ \\
${\bar E_{\gamma}^{obs}}, keV$ & $280$ & $1020$ & $48$ & $125$ \\ \hline
\end{tabular}
\end{center}
\end{table}

\newpage
\small
\begin{table}
\caption{Values of the parameters of hydrodynamically accelerated
matter at the end of the acceleration phase. Numbers in the brackets
for $Q_N$ give the powers of ten.\label{table_3}}
%\begin{center}
\begin{tabular}{ l c c c c  c c c c}
\hline \hline
 & \multicolumn{4}{c}{\em Compact white dwarf} &
\multicolumn{4}{c} {\em Extended white dwarf} \\

 & \multicolumn{2}{c} {$a=0.2$} & \multicolumn{2}{c} {$a=0.232$} &
\multicolumn{2}{c} {$a=0.2$} & \multicolumn{2}{c} {$a=0.232$} \\

Parameters & $b=2.0$ & $b=2.73$ & $b=2.0$ & $b=2.73$ & $b=2.0$ & $b=2.73$ &
$b=2.0$ & $b=2.73$ \\ \hline

$\Gamma_f$ & $45$ & $170$ & $230$ & $1500$ & $12$ & $27$ &
$44$ & $160$ \\

$E^{max}_K, GeV$ & $42$ & $158$ & $214$ & $1400$ & $11$ & $25$ &
$41$ & $150$ \\

$\gamma_{rl}^{int}$ & $3.3$ & $2.4$ & $2.9$ & $2.1$ & $3.3$ & $2.4$ &
$2.9$ & $2.1$ \\

$Q_N(E_K>500MeV)$ & $3.9(48)$ & $3.9(48)$ &
$4.0(49)$ & $4.0(49)$ &
$2.8(48)$ & $2.8(48)$ & $2.9(49)$ & $2.9(49)$\\

$Q_N(E_K>940MeV)$ & $5.2(47)$ & $5.2(47)$ &
$7.0(48)$ & $7.0(48)$ &
$3.7(47)$ & $3.7(47)$ & $5.0(48)$ & $5.0(48)$\\

$Q_N(E_K>9.4GeV)$ & $1.4(43)$ & $1.8(44)$ &
$8.6(44)$ & $7.5(45)$ &
$1.0(43)$ & $1.3(44)$ & $6.1(44)$ & $5.4(45)$ \\ \hline

\end{tabular}
%\end{center}

\newpage
\caption{The parameters of the main gamma ray burst. The numbers in the
brackets for $N_{\gamma}$ give the powers of ten.
\label{table_4}}
%\begin{center}
\begin{tabular}{l c c c c c c c c}
\hline \hline
 & \multicolumn{4}{c}{\em Compact white dwarf} &
\multicolumn{4}{c} {\em Extended white dwarf} \\

 & \multicolumn{2}{c} {$a=0.2$} & \multicolumn{2}{c} {$a=0.232$} &
\multicolumn{2}{c} {$a=0.2$} & \multicolumn{2}{c} {$a=0.232$} \\

Parameters & $b=2.0$ & $b=2.73$ & $b=2.0$ & $b=2.73$ & $b=2.0$ & $b=2.73$ &
$b=2.0$ & $b=2.73$ \\ \hline

$W_{\gamma}, ergs$ & $2.6(42)$ & $6.2(42)$ &
$3.2(43)$ & $7.2(43)$ &
$5.6(41)$ & $1.4(42)$ & $7.0(42)$ &
$1.6(43)$\\

$N_{\gamma}^{tot}, phot$ & $1.5(46)$ & $3.9(46)$ &
$2.0(47)$ & $4.5(47)$ &
$3.5(45)$ & $8.8(45)$ & $4.4(46)$ &
$1.0(47)$\\

$E_\gamma^{m}, GeV$ & $4.2$ & $16$ & $21$ & $140$ & $1.0$ & $2.5$ &
$4.1$ & $15$ \\

$N_{\gamma}(10GeV), ph/GeV$ & $5.0(38)$ & $2.0(40)$ &
$6.0(40)$ & $1.0(42)$ &
$4.0(33)$ & $4.0(37)$ & $8.0(37)$ & $2.0(41)$\\ \hline

\end{tabular}
%\end{center}
\end{table}

\newpage

\figcaption[figura_1.ps]{A schematic model of a presupernova Ia binary system.
The secondary companion (red giant) loses matter via stellar wind and the
white dwarf accretes it in the regime of wind accretion.\label{fig_1}}

\figcaption[figura_2.ps]{The total column density of gas in the binary
system as a function of angle $\theta$ relative to the binary axis
(the line between the centers of two stars).\label{fig_2}}

\figcaption[figura_3.ps]{The time profile of the gamma ray burst.
On the left ordinate axis the
energy-integrated number of photons emitted per $1s$ and $1sr$ is displayed
. The emission time $t$ shown in the lower absciss axis is defined as the
time-of-flight $t=(r-R_S)/v_N$ for the nuclei with energy per nucleon $E_K$
at the maximum of the production rate $<\xi \sigma(E_K)>N(E_K)$. The time
profile curve (full upper curve) coincides with the density
distribution $\rho(r)$ given by the
values on the right ordinate axis and the upper absciss axis. The minimum
distance corresponds to the surface of the star. The curves
are given for the cases of {\it compact} and {\it extended} white dwarfs
and for $\theta=90^o$. Note that the gamma radiation is almost spherically
symmetrical (see Fig. 2).\label{fig_3}}

\figcaption[figura_4.ps]{Energy spectra at $E_{\gamma}\geq 1 GeV$
for {\it compact white dwarf} (solid curves) and {\it extended white dwarf}
(dashed curves). The corresponding values of the parameters $a$ and $b$ are
shown in the
Figure. The plotted quantities are the number of photons
in the burst per $1 GeV$.\label{fig_4}}

\end{document}